\documentclass[aps,preprint]{revtex4}%
\usepackage{amsfonts}
\usepackage{amsmath}
\usepackage{amssymb}
\usepackage{graphicx}%
\begin{document}
\title{Magnetic and superconducting properties of $RuSr_{2}Gd_{1.5}Ce_{0.5}%
Cu_{2}O_{10-\delta}$ samples: dependence on the oxygen content and ageing effects}
\author{C. A. Cardoso}
\affiliation{Departamento de F\'{\i}sica, UFSCar, 13565-905 , Sao Carlos-SP, Brazil}
\author{A. J. C. Lanfredi}
\affiliation{Departamento de F\'{\i}sica, UFSCar, 13565-905 , Sao Carlos-SP, Brazil}
\author{A. J. Chiquito}
\affiliation{Departamento de F\'{\i}sica, UFSCar, 13565-905 , Sao Carlos-SP, Brazil}
\author{V. P. S. Awana}
\affiliation{National Physical Laboratory, K. S. Krishnan Marg., New Delhi 110012, India}
\author{H. Kishan}
\affiliation{National Physical Laboratory, K. S. Krishnan Marg., New Delhi 110012, India}
\author{R. L. de Almeida, O. F. de Lima}
\affiliation{Instituto de F\'{\i}sica \textquotedblleft Gleb Wataghin\textquotedblright,
UNICAMP, 13083-970,Campinas-SP, Brazil}
\author{F. M. Ara\'{u}jo-Moreira}
\affiliation{Departamento de F\'{\i}sica, UFSCar, 13565-905 , Sao Carlos-SP, Brazil}
\begin{abstract}
The magnetic and superconducting properties of $RuSr_{2}Gd_{1.5}Ce_{0.5}%
Cu_{2}O_{10-\delta}$ polycrystalline samples with different oxygen doping
level are presented. A strong suppression of the superconducting temperature
($T_{c}$), as well as a reduction in the superconducting fraction, occurs as
the oxygen content is reduced by annealing the samples in oxygen deprived
atmospheres. Drastic changes in the electrical resistivity are observed above
$T_{c}$, possibly associated with oxygen removal mainly from grain boundaries.
However, the magnetic ordering is relatively less affected by the changes in
oxygen content of the samples. The spin glass transition is enhanced and
shifted to higher temperatures with the reduction in oxygen content. This
could be correlated with an increase in the spins disorder and frustration for
the oxygen depleted samples. Also, the same oxygen vacancy induced disorder
could explain the reduction in the fraction of the sample showing
antiferromagnetic order. We also report significant changes in the measured
properties of the samples as a function of time.


\end{abstract}
\maketitle

\section{Introduction}

The rutheno-cuprate $RuSr_{2}Gd_{1.5}Ce_{0.5}Cu_{2}O_{10-\delta}$ (Ru-1222)
and $RuSr_{2}GdCu_{2}O_{8-\delta}$ (Ru-1212) families have attracted
considerable interest by the coexistence of magnetic order and
superconductivity in these compounds \cite{Felner97,Bernhard99}. The magnetic
ordering of the Ru moments occurs at a temperature around 100 K, while the
superconducting transition temperature ($T_{c}$) is usually no higher than 50
K. The fact that superconductivity occurs in the $CuO_{2}$ planes, while the
magnetic order is related with the \textsl{Ru} ions, casts some doubt about
the genuine coexistence of these two phenomena at a microscopic level. Also,
the exact nature of the magnetic order is still in debate. In this context,
the role played by the oxygen non-stoichiometry in the determination of the
magnetic and superconducting properties of these compounds is still not
completely clear \cite{reviewAwana,Matvejeff,Cardoso04}. In particular, the
large number of oxygen vacancies observed in Ru-1222, bigger than in Ru-1212
\cite{Matvejeff}, was associated with the presence of a spin glass (SG)
behavior observed in Ru-1222 samples \cite{Cardoso04,Cardoso03}. Although
reports on the oxygen doping in Ru-1222 can be found in the literature
\cite{Shi,Felner01,Felner02,Felner00}, the possible correlation of oxygen
stoichiometry and the appearance of the SG phase was not explored yet. In
fact, this has been one of the motivations for the present work. We have
studied polycrystalline samples of Ru-1222 with different oxygen doping levels
by performing magnetization, ac susceptibility and resistivity measurements.
For oxygen depleted samples we observed a large variation in the resistivity
and a suppression of the superconducting transition, while the magnetic order,
above $T_{c}$, showed a subtle qualitative modification. A significant
variation in the measured properties of the samples was also observed as a
function of time, which may help to explain some contradictory results
reported in the literature.

\section{Experimental Details}

Samples of composition $RuSr_{2}Gd_{1.5}Ce_{0.5}Cu_{2}O_{10-\delta}$ were
synthesized through a solid-state reaction route. High purity $RuO_{2}$,
$SrCO_{3}$, $Gd_{2}O_{3}$, $CeO_{2}$ and $CuO$ were mixed in the
stoichiometric proportions and calcinated at 1000, 1020, 1040 and 1050 $^{o}C$
each for 24 hours with intermediate grindings. One pellet was annealed in a
flow of high pressure oxygen (100 atm) at 420 $^{o}C$ for 100 hours and
labeled as O$_{2}$-annealed. Two other pellets were annealed in air for 48
hours at 1050 $^{o}C$ (labeled as air-annealed). One of them was further
annealed in nitrogen at 420 $^{o}C$ for 24 hours (N$_{2}$-annealed). In this
way it was possible to vary the oxygen content in the studied samples. Though
the actual oxygen content of the samples were not determined, qualitatively
the same should be maximum for 100 atm O$_{2}$ annealed, moderate for air
annealed and minimum for the N$_{2}$-annealed sample.

X-ray diffraction (XRD) patterns were obtained at room temperature for all
three samples (MAC Science: MXP18VAHF$^{22}$; Cu\textsl{K}$_{\alpha}$
radiation) . The ac susceptibility measurements were performed in a commercial
PPMS (Physical Properties Measurement System), while for the dc measurements a
SQUID magnetometer MPMS-5 was employed, both equipments made by Quantum Design
company. For the resistivity measurements the four-point technique was used.

\section{Results and discussion}

Analysis of the XRD patterns (Fig. 1) revealed that all three samples
crystallize in a tetragonal structure (space group \textsl{I4/mmm}) with the
following lattice parameters for O$_{2}$-, Air- and N$_{2}$-annealed samples,
respectively: $a=b=3.8327(7)$ \AA \ and $c=28.3926(8)$ \AA ; $a=b=3.8427(7)$
\AA \ and $c=28.4126(8)$ \AA ; and $a=b=3.8498(3)$ \AA \ and $c=28.4926(9)$
\AA . The increase in the lattice parameters from the O$_{2}$- to the N$_{2}%
$-annealed samples indicates that the different annealings are not only
affecting the grain boundaries but leading to an overall reduction in the
oxygen content in the bulk of the air- and N$_{2}$-annealed samples. Small
impurity peaks were observed in the XRD patterns, which are the same for all
samples. Despite the presence of these impurity peaks, our samples may be
considered to be of very good quality if compared to those reported in the
literature by several authors \cite{Awana04}.

The temperature dependence of the magnetization for all three samples is
presented in Fig. 2. The samples were cooled in zero magnetic field down to
the lowest accessible temperature (2 K). After temperature stabilization, a
magnetic field of 50 Oe was applied and the magnetization recorded as the
temperature was raised (ZFC curve) up to 200 K. Then the temperature was
decreased back to 2 K, keeping the same applied magnetic field, and the FC
curve was obtained. The ZFC curves show a pronounced peak at 68 K, 84 K and 92
K for O$_{2}$-, Air- and N$_{2}$-annealed samples, respectively, while the FC
curves show a monotonical increase with the reduction of the temperature.
Besides the shift to higher temperatures of the peak in the ZFC curve, its
height also increase with the reduction in the oxygen content of the samples,
going from 0.58 emu/g for the O$_{2}$- annealed sample to 0.67 emu/g for
N$_{2}$- annealed one. In the SG scenario, the frustration of the
antiferromagnetic interaction is believed to be associated with the presence
of disorder due to the large number of oxygen vacancies in this compound
\cite{Matvejeff,Cardoso03}. Thus, an increase in the number of oxygen
vacancies would lead to a more disordered system, favoring the occurrence of
the SG phase at higher temperatures. This interpretation is consistent with
the results presented in Fig. 2. A small antiferromagnetic (AFM) transition
can be detected at higher temperatures, around 174 K for all samples, although
this transition becomes less pronounced with the reduction in oxygen content.
At temperatures between the AFM transition ($\sim174$ K) and the SG freezing
point ($\sim80$ K) a small irreversibility can be observed. As shown in the
insets of Fig. 2, the irreversibility is greatly reduced for the air-annealed
sample and it is almost completely absent in the N$_{2}$-annealed sample.
These results point to the simultaneous occurrence of an ordered phase (AFM)
and a disordered or frustrated phase \cite{Yoshizawa}, whose configurations
freeze into a SG at Tf $\sim$80 K. It is likely that as oxygen vacancies are
introduced in the sample, a larger fraction of the same becomes disordered
favoring the SG phase at the expenses of the AFM phase. In fact, the almost
reversible magnetization curve observed for the N$_{2}$-annealed sample, above
the SG freezing temperature $T_{f}$, exhibits the expected behavior of a
prevailing SG phase \cite{Binder}.

Fig. 3 presents the temperature dependence of the real part of the complex ac
susceptibility $\chi_{ac}=\chi^{\prime}+i\chi^{\prime\prime}$, measured with
the same routines employed to obtain the magnetization curves. The
measurements were performed at an applied field of $H=50$ Oe and revealed well
defined peaks at temperatures $T_{f}=70.7$ K$,$ $87.9$ K and $93.7$ K for
O$_{2}$-, Air- and N$_{2}$-annealed samples, respectively. The position of
these peaks defines the freezing temperature of the spin system, where the ZFC
and FC curves branch apart from each other in the magnetization measurements
(see Fig. 2). For the ac susceptibility, the ZFC and FC curves are almost
identical for all three samples, except for temperatures in the 40 - 90 K
range. For these temperatures a small irreversibility is observed for all
samples, being more pronounced for the air-annealed sample and less pronounced
for the O$_{2}$-annealed sample. A clear feature present in these curves is a
sudden decrease in $\chi^{\prime}$, associated with the superconducting
transition. We see that the superconducting phase is strongly suppressed by
the reduction of the oxygen content in the samples. Both the transition
temperature and the superconducting fraction are reduced until the complete
suppression for the N$_{2}$-annealed samples, which does not present
superconductivity down to 2 K.

The frequency dependence of the peak in $\chi^{\prime}$ as a function of the
temperature is a clear signature of a SG phase. To check this we repeated the
$\chi^{\prime}\times T$ measurements for three different frequencies, for all
samples. For the sake of brevity, we show in Fig. 4 only the results for the
N$_{2}$-annealed sample. Similar results were found also for the other
samples. As observed in Fig. 4, the peak shifts to lower temperatures and its
intensity increases as the frequency of the excitation field is decreased,
which is the expected behavior for a SG system
\cite{Livro,Cardoso03,Cardoso04}. The results obtained through magnetization
and ac susceptibility measurements show unequivocally the presence of a SG
phase in all three samples studied, being more noticeable in the oxygen
depleted ones. On the other hand, a reduction in the oxygen content suppresses
both the superconducting phase (below $T_{f}$) and the fraction of the sample
that presents AFM order (above $T_{f}$).

The imaginary component of the complex susceptibility also presents a peak
associated with the glassy transition, as can be observed in Fig. 5. This peak
occurs about 3 K below the corresponding peak in $\chi^{\prime}$ and is weakly
frequency-dependent, as expected for a SG phase \cite{Binder}. However, a
second peak appears at 42.9 K, 57.7 K and 65.7 K for the O$_{2}$-, air- and
N$_{2}$-annealed samples, respectively, as shown in Fig. 5 for a frequency of
the driving field of 1000 Hz. This peak is strongly frequency-dependent, being
more intense and shifted to higher temperatures as the frequency is increased,
as shown for the N$_{2}$-annealed sample in Fig. 6. It is important to notice
that the strong frequency dependence of this second peak contrasts with the
very weak frequency dependence observed for the SG peak, clearly indicating
that it has a different origin. It shows a more intense signal in the FC
branch, if compared with the ZFC branch, thus revealing a strong
irreversibility. Also, the temperature range where this peak and the
irreversibility appear in the $\chi^{\prime\prime}$ measurements coincide with
the irreversibility observed in $\chi^{\prime}$ (see Fig. 3). For the O$_{2}$
sample this peak could tentatively be associated with the superconducting
transition, which occurs at a close temperature. However, the same is not true
for the other two samples. For the air-annealed sample, for instance, the peak
related to the superconducting transition is located about 47 K below the
second peak position. A more likely explanation for the irreversibility and
the second peak in $\chi^{\prime\prime}$ could be associated with the rotation
of spins or spin-clusters, due to the excitation field applied during the ac
susceptibility measurements. A recent report on samples of the family Ru-1222
shows \cite{Felner04} that for temperatures in between 30 and 80 K, depending
of the exact sample composition, the hysteresis loops are very narrow, with
the coercivity field ($H_{c}$) being zero at a certain temperature within this
interval. This coincides very well with the temperature range where we found
an irreversible behavior in our $\chi^{\prime\prime}$ measurements. If $H_{c}$
is small enough, then the amplitude of our ac magnetic field (1 Oe) could
rotate the existing magnetic domains. Since this spin rotation is a
dissipative process, one should expect an increase in the dissipation (and
then in $\chi^{\prime\prime}$) which could peak when $H_{c}=0$. Also, it was
shown that the coercive field is larger for samples annealed in high pressure
of oxygen, compared with as grown samples \cite{Felner00}. Therefore it can be
inferred that this peak in $\chi^{\prime\prime}$ should be stronger for oxygen
depleted samples, in full agreement with our data (see Fig. 5). Although no
clear change at this temperature can be detected in the magnetization curves
(Fig. 1) for any of the three samples, it is possible to argue that this peak
indicates a re-arrangement of the spins in the sample, going from a more
disordered state (SG phase) to a more ordered weak-ferromagnetic (W-FM) state.
Our results alone are insufficient to clearly identify such re-ordering of the
spins, however recent results on zero field muon spin rotation (ZF-$\mu$SR)
have shown two different internal fields at low temperatures, one vanishing at
temperatures above 90 K and the second one at 80 K \cite{Shengelaya}. 

Another important property strongly affected by the change in the oxygen
content of the samples is their resistivity, as shown in Fig. 7. By reducing
the oxygen content of the sample both the resistivity and its variation with
temperature increase (see Fig. 7(a)). All samples present an exponential
growth of resistivity with decreasing temperature. At low temperatures the
behavior changes: for the O$_{2}$ and air annealed samples, the
superconducting transition can be undoubtedly identified, while for the
N$_{2}$ samples the resistivity continues to increase at an even higher rate.
A different result was obtained when we repeated these measurements after a
few months. A clear increase in the resistivity was observed, indicating a
possible continuous loss of oxygen by the samples, which were kept in a
dessecator all the time. After 5 months (see Fig. 7(b)), the resistivity
measurements for the air-annealed sample (open squares) is similar to the
result previously obtained for the N$_{2}$ sample, showing no sign of
superconductivity. To evaluate if this sample degradation could be due to
oxygen loss only, we tried to recover the original state of the air-annealed
sample by annealing it in an oxygen atmosphere. Our attempts to re-oxygenate
the sample in a 1 atm of oxygen atmosphere for 100 hours at 450 $^{o}$C and
100 hours at 800 $^{o}$C did not produce any meaningful change in its
properties. Then it was annealed again for 100 hours at 420 $^{o}$C in 70 atm
of O$_{2}$. The resistivity curve obtained after oxygenation, shown in Fig
7(b) as crossed squares, presented a drastic reduction in its slope, becoming
similar to the curve for the O$_{2}$-annealed sample. Nevertheless, the
superconductivity could not be restored at all. After a second annealing,
performed under the same conditions as the previous one, the superconducting
transition could then be partially recovered in this sample (Fig 7(b), solid
squares). The obtained $T_{c}\approx25$ K indicates that even after both
annealings the sample still remains underdoped. Interestingly, the normal
state resistivity for the sample after both annealings processes shows a
steeper change in resistivity with the decrease in temperature. An $M\times T$
curve for $H=50$ Oe (ZFC followed by FC) was also measured for the
re-oxygenated sample as shown in Fig. 8. We can see that the basic features
previously observed for the oxygen annealed sample are all present, including
the irreversibility observed at temperatures immediately above the onset of
the SG peak and the AFM transition at $T_{c}\approx175$ K (see inset of Fig.
8). Both resistivity and magnetic measurements indicate that the oxygen
content in the sample increased with the annealings but they were not enough
to fully re-oxygenate the sample. The fact that the resistivity is strongly
affected by the degradation of the sample and by its re-oxygenation, while the
changes in magnetization and susceptibility are more subtle, seems to indicate
that both processes affect primarily the grain boundaries. However, the
complete suppression of the superconductivity, not detected even by inductive
measurements, and the change in the AFM transition for the degraded samples,
point to the relevance of processes occurring in the bulk of the grains.

Our results indicate that although the oxygen can easily leave the sample, the
reverse process of re-oxygenation is quite tricky and difficult. Also, the
oxygen content of Ru-1222 samples can change with time and that ageing process
could be the origin for many discrepancies between results reported in the
literature. For instance, the valence of the ruthenium ions is known to be
affected by the change in oxygen content \cite{reviewAwana}. Our results
indicate that not only the synthesis process may be responsible for the
different values of Ru valence reported in the literature, due to the
different initial oxygen content of the samples. The time delay between the
samples preparation and their measurements could also strongly affect the results.

\section{Discussion}

The magnetic properties of Ru-1222 are very complex, with different magnetic
transitions and possible phase separation. The significant role played by
oxygen vacancies is still under study and would be benefited by a careful
comparison of works from different groups. It is important to compare results
obtained by different techniques in an attempt to reach a better understanding
on the interesting properties of this rutheno-cuprate family. Recently, Felner
et al. \cite{Felner04} presented a detailed M\"{o}ssbauer study of Ru-1222
where they observed two magnetic transitions. The first was an AFM transition
at 160 K, and the second was identified as the onset of a W-FM order at around
90 K. Interestingly, only 10-20\% of the sample volume orders
antiferromagnetically while its major fraction presents a magnetic order only
below 90 K. These results gave strong support to the idea of magnetic phases
separation in Ru-1222 polycrystalline samples \cite{Xue}. Further, the
M\"{o}ssbauer spectra was rather broad and more than one sub-spectra, besides
the one associated with the small AFM fraction, was necessary to fit the
experimental results, indicating some degree of disorder or inhomogeneity in
the sample. Both central conclusions presented in that work, the occurrence of
phase separation and the presence of disorder/inhomogeinity are consistent
with our results. Similar conclusions were reached by Shengelaya et al.
\cite{Shengelaya} based on muon spin rotation (%
$\mu$%
SR) experiments in $RuSr_{2}Eu_{1.4}Ce_{0.6}Cu_{2}O_{10}$ samples. They
observed that a fraction around 15\% of the sample orders magnetically at
temperatures around 200 K, while the remaining fraction of the sample orders
only at temperatures below 90 K, in close connection with the strong increase
in magnetization. Then, a well resolved internal field appears at a lower
temperature close to 77 K. Another important result is that two different
oscillating components were observed. This may indicate either two different
muon stopping sites within the same magnetic phase or two different magnetic
phases, with two different but close temperature transitions. Comparing these
results with ours, we find again support for the idea of phase separation,
with a possible occurrence of AFM clusters at temperatures between 90 and 200
K. The majority phase orders at a temperature of 77.6 K, which is close to the
freezing temperature observed in our O$_{2}$-annealed sample. Our results have
demonstrated a strong sensitivity of the magnetic properties of Ru-1222 to the
sample synthesis process and oxygen content, that changes dramatically from
sample to sample and, for the same sample, with time. However, some common
ground has been achieved. It seems clear that two different magnetic phases
are present in Ru-1222: one AFM phase, that shows a transition temperature
around 160 -- 200 K, and another phase that appears at temperatures around 70
-- 90 K. The AFM phase presents spin canting that leads to a small
ferromagnetic loop, and it is usually a minority phase corresponding to 10 --
20 \% of the sample volume. Our results indicate that this fraction decreases
with the reduction of the oxygen content. The second magnetic transition
occurs at $\sim90$ K when the rest of the sample starts to show some magnetic
order. The frequency dependence observed in our ac susceptibility data
strongly suggests that a SG phase sets in at this temperature leading to an
additional ferromagnetic component. As the oxygen content is reduced and the
system becomes more disordered (or more inhomogeneous) this SG phase is
favored and both its volume fraction and freezing temperature increases. The
dissipative peak observed at temperatures around 60 K (see Fig. 5), in the
imaginary component of the ac susceptibility, seems to be related with a
coercive field being close to zero in this temperature range
\cite{Felner00,Felner04}. This fact may indicate some reordering of the spins,
from the SG phase into a more ordered W-FM state. However, no significant
change was observed in the magnetization curves, so if such reordering really
occurs it should be quite subtle.

\section{Final remarks}

In this work we have explored the influence of the oxygen content in the
electrical and magnetic properties of polycrystalline samples of
$RuSr_{2}Gd_{1.5}Ce_{0.5}Cu_{2}O_{10-\delta}$. It was observed that the
reduction in the oxygen content favors the development of a spin glass state
at the expenses of the antiferromagnetic phase. The presence of oxygen
vacancies in the system may affect the magnetic ordering of the Ru sublattice
in two ways: by changing the valence of some of the \textit{Ru} ions and by
distorting the crystalline lattice. Both mechanisms introduce disorder in the
system and affect the antiferromagnetic interaction, favoring the appearance
of a spin glass phase. By reducing the charge density in the system, the
oxygen depletion also strongly influences the resistivity and the occurrence
of superconductivity in Ru-1222, which disappears completely for the N$_{2}%
$-annealed sample.

We also documented the deterioration of the samples in a time window of a few
months and its recovery after annealing in a high pressure atmosphere of
oxygen. In a polycrystalline sample it is expected that oxygen first leaves
the grain boundaries, thus strongly affecting the resistivity of the sample.
Further reduction in the oxygen content occurs at the $RuO_{2}$ planes
\cite{Matvejeff}, affecting strongly the magnetic order and leading to the
spin glass phase. In a later stage the oxygen from the $CuO_{2}$ planes
possibly starts to leave the sample suppressing gradually the
superconductivity. Similarly, in the re-oxygenation process the grain
boundaries are easily oxygenated, leading to a quick recover in the
resistivity of the sample. However, the oxygen diffusion inside the grains is
difficult and slow, being effective only when the sample is annealed in a high
oxygen pressure. Since this process was incomplete in our samples the spin
glass phase persisted even after a high oxygen filling. By further increasing
the oxygen content it starts to get into the $CuO_{2}$ planes and then
superconductivity is recovered. It seems that some rearrangement of the oxygen
atoms inside the sample may take place at this point, what could eventually
explain why the recovery of superconductivity occurs simultaneously with an
increase in the normal state resistivity as observed in this work. In
conclusion, these results corroborate previous discussions on the importance
of the oxygen doping level in the electrical transport and magnetic properties
of Ru-1222 samples \cite{Cardoso04,Cardoso03}. Additionally, we have found
that ageing effects could be a possible explanation for some of the
contradictory results found in the literature for this ruthenocuprate family.

This work was partially supported by Brazilian agencies FAPESP and CNPq.
\bigskip

\newpage

\begin{center}
FIGURE CAPTIONS
\end{center}

Figure 1 - X-ray diffraction patterns for all three samples studied.

\bigskip

Figure 2 - Magnetization as a function of temperature and $H=50$ Oe, for all
three samples. Insets show the suppression of the irreversibility observed
near the onset of the magnetic transition, with decreasing the oxygen content.

\bigskip

Figure 3 - Real part of the ac magnetic susceptibility as a function of
temperature at an applied dc field of $H=50$ Oe, for all three samples. The
amplitude and frequency of the excitation field were 1 Oe and 1000 Hz, respectively.

\bigskip

Figure 4 - Real part of the ac susceptibility as a function of temperature for
$H=50$ Oe and frequencies of 100, 1000 and 10000 Hz, for the N$_{2}$-annealed sample.

\bigskip

Figure 5 - Imaginary part of the ac magnetic susceptibility as a function of
temperature at an applied dc field of $H=50$ Oe, for all three samples. The
amplitude and frequency of the excitation field were 1 Oe and 1000 Hz, respectively.

\bigskip

Figure 6 - Imaginary part of the ac susceptibility as a function of
temperature for $H=50$ Oe and frequencies of 100, 1000 and 10000 Hz, for the
N$_{2}$-annealed sample. The data for 1000 Hz and 10000 Hz were respectively
shifted up by 1.0 and 2.0, in units of the arbitrary scale, for a better visualization.

\bigskip

Figure 7 - (a) Normalized resistivity as a function of temperature, for all
three samples. (b) results for the air-annealed sample after degradation and
re-annealing in O$_{2}$ atmosphere (see text).

\bigskip

Figure 8 - Magnetization as a function of temperature and $H=50$ Oe, for the
air-annealed sample after degradation and re-annealing. The inset shows the
irreversibility near the onset of the magnetic transition.

\end{document}